\documentstyle[12pt,epsf]{article}
\catcode`@=11  
\def\@refe#1{#1}
\def\@biblabel#1{{\small\bf{#1}}}
\def\refe{\@ifnextchar [{\@tempswatrue\@citexr}{\@tempswafalse\@citexr[]}}
\def\@citexr[#1]#2{\if@filesw\immediate\write\@auxout{\string\citation{#2}}\fi
  \def\@citea{}\@refe{\@for\@citeb:=#2\do
    {\@citea\def\@citea{,}\@ifundefined
       {b@\@citeb}{{\bf ?}\@warning
       {Citation `\@citeb' on page \thepage \space undefined}}%
\hbox{\csname b@\@citeb\endcsname}}}{#1}}
 \catcode`@=12

\def\figdir{}

\def\autofig#1#2#3#4#5{\begin{figure}[#1]\begin{center}
\epsfxsize=#2\textwidth\leavevmode\epsffile{\figdir #3.fig}\end{center}
\caption[fake]{{\footnotesize #4}\label{#5}}
\end{figure}}        
\def\getFIG#1#2{\epsfxsize=#1\textwidth\leavevmode\epsffile{\figdir #2.fig}}

\catcode`\@=11 

\def\mypart{\clearpage
   \thispagestyle{plain}%
   \global\@topnum\z@
   \@afterindentfalse
   \secdef\@mypart\@schapter}

\def\@mypart[#1]#2{\ifnum \c@secnumdepth >\m@ne
        \refstepcounter{chapter}%
        \typeout{\@chapapp\space\thechapter.}%
      \addcontentsline{toc}{chapter}{#1}\fi
    \markboth{\ifnum \c@secnumdepth >\m@ne
                  #1  \fi}{}%
   \addtocontents{lof}%
       {\protect\addvspace{10\p@}}
   \addtocontents{lot}%
       {\protect\addvspace{10\p@}}
   \if@twocolumn
           \@topnewpage[\@makemyparthead{#2}]%
     \else \@makemyparthead{#2}%
           \@afterheading
     \fi}
\def\@makemyparthead#1{%
  \vspace*{50\p@}%
  {\parindent \z@ \raggedright
    \ifnum \c@secnumdepth >\m@ne
      \huge\bf #1
      \par
      \vskip 20\p@ \fi
    \nobreak
    \vskip 40\p@
  }}

\def\defi{=}  
\def\^{}    
\def\iden{\^ 1}   
\def\<{\langle} 
\def\>{\rangle} 

\def\de{\partial}
\def\al{\alpha}
\def\init{\begin{eqnarray}}
\def\fin{\end{eqnarray}}
\def\2righe{\vskip 2 \baselineskip}
\def\riga {\vskip 1 \baselineskip}
\def\meq{master equation}
\def\eq{e\-qua\-tion}

\def\tr{\mbox{Tr}}

\def\gort{\gamma_\perp}
\def\gpar{\gamma_\|}

\def\matd{{\vec{\vec D}}}

\def\Begref{\subsection*{\normalsize References}\begin{description}
\itemsep=0pt\listparindent=0pt\ft }

\def\Endref{\end{description}}
\newcounter{bb}
\def\bib#1{\stepcounter{bb}\bibitem[{\footnotesize\thebb}]{#1}}
\def\<{\langle}\def\>{\rangle}
\def\pni{\par\noindent}
\def\ft{\footnotesize}

\def\bold#1{\setbox0=\hbox{$#1$}%
     \kern-.025em\copy0\kern-\wd0
     \kern.05em\copy0\kern-\wd0
     \kern-.025em\raise.0433em\box0 }

\def\f{\frac}

\def\al{\alpha}
\def\alb{{{\alpha}^*}}
\def\ad{a^{\dag}}

\def\dt{\partial_t}

\def\dx{\partial_x}
\def\ddx{\partial^2_{xx}}

\begin{document}
\small
\title{\large\bf Noise, errors and information in quantum amplification}
\author{\normalsize G.M. D'Ariano$^a$, C. Macchiavello$^b$ and 
L. Maccone$^a$. \\ 
{\normalsize\em a) Dipartimento di Fisica \lq\lq A. Volta\rq\rq, via Bassi
6, I27100 Pavia, Italy}\\
{\normalsize\em b) Clarendon Laboratory, Oxford University, OX1 3PU Oxford, UK}}
\date{}
\maketitle
\small
\begin{abstract}
We analyze and compare the characterization of a quantum device in
terms of noise, transmitted bit-error-rate (BER) and mutual
information, showing how the noise description is meaningful only for
Gaussian channels.  After reviewing the description of a quantum
communication channel, we study the insertion of an amplifier. We
focus attention on the case of direct detection, where the linear
amplifier has a 3 decibels noise figure, which is usually considered
an unsurpassable limit, referred to as the standard quantum limit
(SQL). Both noise and BER could be reduced using an ideal amplifier,
which is feasible in principle.  However, just a reduction of noise
beyond the SQL does not generally correspond to an improvement of the
BER or of the mutual information.  This is the case of a laser
amplifier, where saturation can greatly reduce the noise figure,
although there is no corresponding improvement of the BER.  Such
mechanism is illustrated on the basis of Monte Carlo simulations.
\end{abstract}

\section{Introduction}
\label{int}

In order to exploit the bandwidth available in the optical domain the
evolution of optical communications urges conversion of hybrid
electro-optical devices towards all-optical ones. For long distance
communications the losses along the optical fiber decrease the
transmitted power and introduce communication errors, thus being the
crucial limitation to the development of this kind of technology.  On
the other hand, an amplifier along the line restores the power level,
but introduces noise of quantum origin. For direct detection, which
would allow to achieve the ultimate channel capacity \cite{OY}, an
optical phase insensitive amplifier used in the linear regime
introduces 3 decibels of noise. Such noise has long been considered as
an unsurpassable limit \cite{3db} --- the so called ``standard quantum
limit'' (SQL)--- which, however is just a peculiarity of the linear
phase insensitive amplifier (PIA) \cite{saturation}.

In order to recover the transparency of an optical network one needs
to achieve a perfect amplification that, by itself, does not introduce
any additional disturbance.  Indeed, as suggested by Yuen
\cite{Yuen86}, it is possible in principle to achieve such ideal
amplification also for direct detection, although there is still no
actual device that can attain it in practice.  With this aim one could
try to improve the performance of an existing amplifier by driving it
far from the linear regime, and going beyond the SQL.  In
Ref. \cite{saturation} it has been shown that this is indeed possible
for a saturable laser amplifier, where in an intermediate-saturating
regime one can accomplish noise suppression with still sizeable gains.  The
objective of Ref. \cite{saturation} was originally to prove that the SQL can
be actually breached in a concrete case. In a following debate
\cite{comment,reply} it has been pointed out that the noise is not the
significant quantity for evaluating the goodness of an amplifier, and
one should better resort to the original problem of the transmitted
bit-error-rate (BER). However, Ref. \cite{comment} still considered the
PIA as the reference standard quantum amplifier, whereas, as clarified
in Ref. \cite{reply}, the ideal photon number amplifier (PNA) of Yuen
indeed could greatly improve the transmitted BER of the PIA.

In this paper we reconsider the problem of characterizing the quality
of a low-noise amplifier from the beginning. We clarify that the
quantity that unambiguously determines the behavior of a device
inserted in a communication channel is the so called ``mutual
information'' between the input and the output.  For a binary channel,
the BER is essentially equivalent to the mutual information for small
error probabilities, whereas it remains an ambiguous characterization
for any pathological situation.  On the other hand, we will also show
that the characterization in terms of noise is still meaningful for
Gaussian communications channels.

As the noise characterization \cite{saturation} is not sufficient to
establish whether a saturable amplifier can perform better than a
linear one, here we present a careful analysis of a laser amplifier in
a quantum regime, and compare it with an ideal PIA.  We illustrate the
mechanism that underlies a noise reduction under saturation, and
explain why such reduction does not correspond to any improvement of
the BER.  We will show that this is typical of the saturation
mechanism, whereas, in general, a noise reduction may correspond to an
improvement of the BER, and, indeed, for an ideal PNA such an
improvement can be actually achieved.

The outline of the paper is the following.  In Sec. \ref{basics} we
give a brief introduction to the basic concepts of information theory
that are needed for a complete characterization of the operation of a
quantum amplifier. Such concepts are then specialized to the case of a
quantum communication channel in Sec. \ref{qcc}, where a complete
description of the transmission of quantum signals and the
characterization of a quantum amplifier are given.  In Sec.
\ref{gauss} the linear Gaussian channel is discussed.  This channel is
of particular interest because it is the only case where a quantum
amplifier can be equivalently characterized in terms of mutual
information and in terms of noise.  In view of the subsequent
comparison between the performances of a linear and a saturable
amplifier, in Sec. \ref{s:PIA} the conventional linear PIA is
reviewed and the SQL is derived. In Sec. \ref{s:saturable} the
laser saturable amplifier is presented and analyzed on the basis of
the Fokker-Planck equation derived by Haake and Lewenstein
\cite{haake}. In Sec. \ref{checks}, after presenting some numerical
tests of the theory and of the Monte Carlo simulation
method, we compare the performance of the saturable amplifier to the
one of the linear PIA.  The conclusions are then given in Sec.
\ref{conclusions}.

\section{Basic concepts of information theory}
\label{basics}

A communication channel between a transmitter and a receiver can be
viewed as composed of three basic constituents: an encoder, a
transmission line, and a decoder. The encoder transforms the input of
the channel ---i.e. the message to be transmitted--- into a proper set
of physical signals that are impinged into the transmission line.  In
this schematic representation the transmission line includes all
possible physical processes that affect the signal that carries the
message from the sender to the receiver, including all noise sources
and any kind of device (e.g. amplifiers) inserted along the line.  The
decoder applies a set of operations on the received signal, like
measurements and data processing of the results, and then gives the
reconstructed message at the output of the channel.  If no
transformation modifies the signals along the line, then the encoded
message can be perfectly reconstructed, otherwise errors can arise at
the decoder, and the intrinsic potential of the channel decreases.

The significant quantity that measures the efficiency of a
communication channel is the mutual information. Let us define it
first in the case of a classical discrete channel, and then generalize
it later to the continuous and quantum cases. 

By definition, a discrete channel transmits only symbols belonging to
a numerable set. Suppose that the input of the channel is a symbol
from the set (or alphabet) $\{x_j, j=1,...,J\}$ of $J$ elements with
{\it a priori} probabilities $\{p_j, j=1,...,J\}$ ($\sum_{i=1}^J
p_i=1$), whereas the output is a symbol from the set $\{y_k,
k=1,...,K\}$ of $K$ elements (the two sets need not be equal). The two
sets are linked by specifying the conditional probabilities $Q_{k|j}$
that the output is $y_k$ given the symbol $x_j$ at the input.
The mutual information is given by
\begin{eqnarray}
I(X;Y)=\sum_{j=1}^J\sum_{k=1}^K p_j Q_{k|j}\log_2 \frac{Q_{k|j}}{\sum_i p_i
Q_{k|i}} 
\label{mutinf}
\end{eqnarray}
and quantifies the degree of knowledge (in bits) that the output
random variable $Y$ gives about the input random variable $X$ (by the
capital letter $X\equiv\{x_j, p_j,j=1,\cdots,J\}$ we denote the random
variable, namely the set of symbols $\{x_i\}$ along with the
corresponding probability distribution $\{p_i\}$).  The mutual
information can also be written as follows
\begin{eqnarray}
I(X;Y)=H(X)-H(X|Y)\;,
\label{diff}
\end{eqnarray}
namely as the difference between the entropy at the input
\begin{eqnarray}
H(X)=-\sum_{j=1}^J p_j \log_2 p_j
\label{inp-ent}
\end{eqnarray}
and the conditional entropy at the output
\begin{eqnarray}
H(X|Y)=-\sum_{k=1}^K q_k \sum_{j=1}^J P_{j|k}\log_2 P_{j|k}\;,
\label{cond-ent}
\end{eqnarray}
where $q_k=\sum_i p_i Q_{k|i}$ is the unconditioned probability of the
output symbol $y_k$ and $P_{j|k}=p_jQ_{k|j}/q_k$ is the conditioned
probability that symbol $x_j$ was transmitted given that symbol $y_k$
is received. In the limiting case $X\equiv Y$ that input and output
alphabets coincide and the channel is noiseless
($Q_{j|k}=\delta_{jk}$), one has
\begin{eqnarray}
I(X;X)=H(X)\;,
\label{IXX}
\end{eqnarray}
namely the information transmitted through a noiseless channel is just
the entropy of the input alphabet. In the opposite limiting case that
the input and output random variables are completely uncorrelated
($Q_{j|k}=q_j$ for all $k$) the mutual information vanishes.

The definition of mutual information (\ref{mutinf}) can be
straightforwardly generalized to the case of a continuous set of
symbols $x\in {\cal X}$ and $y\in {\cal Y}$, with {\it a priori}
probability density $p(x)$ and conditional probability density
$Q(y|x)$. Here the sums in (\ref{mutinf}) replaced by integrals as
follows
\begin{eqnarray}
I(X;Y)=\int_{\cal X} dx \int_{\cal Y} dy\ p(x) Q(y|x) 
\log_2 \frac{Q(y|x)}{\int dz\ p(z)Q(y|z)}\;.
\label{minf-cont}
\end{eqnarray}

Let us now consider the case of a binary channel, where two different
symbols ``0'' and ``1'' (which make one ``bit'') are transmitted with
equal {\it a priori} probabilities $p_0=p_1=1/2$.  Usually a binary
channel is characterized by the probability of making errors at the
decoder, called ``bit error rate'' (BER), and defined as follows
\begin{eqnarray}
B=\frac{1}{2}\Bigl(Q_{0|1}+Q_{1|0}\Bigr)\;.
\label{BER}
\end{eqnarray}
In Eq. (\ref{BER}) $Q_{1|0}$ is named ``false alarm probability'' (the
probability of detecting ``1'' when the signal ``0'' was transmitted),
whereas $Q_{1|1}=1-Q_{0|1}$ is called ``detection probability''.
Among the possible choices of sets of independent probabilities, in
the following we adopt the couple $\{Q_{0|1},Q_{1|0}\}$, in terms of
which all quantities of interest (e.g. BER $B$ and mutual information
$I$) can be expressed. The mutual information of the binary channel
takes the form
\begin{eqnarray}
I&=&\frac{1}{2}\Biggl[(1-Q_{1|0})\log_2 \frac{2(1-Q_{1|0})}{1-Q_{1|0}+Q_{0|1}} 
+Q_{0|1}\log_2 \frac{2Q_{0|1}}{1-Q_{1|0}+Q_{0|1}}+\cr &+&
Q_{1|0}\log_2 \frac{2Q_{1|0}}{1-Q_{0|1}+Q_{1|0}} +
(1-Q_{0|1})\log_2 \frac{2(1-Q_{0|1})}{1-Q_{0|1}+Q_{1|0}}\Biggr]\;. 
\label{IB}
\end{eqnarray}
In the binary channel the BER gives the same characterization as the
mutual information in the limiting case of small error probabilities
($Q_{1|0},Q_{0|1}\ll 1$). Here, $I$ can be expanded at first order in
$Q_{1|0}$ and $Q_{0|1}$, and one has
\begin{eqnarray}
B\ll 1 \Rightarrow I\cong 1-B\;.
\label{equi}
\end{eqnarray}
Notice, however, that the BER and the mutual information generally do
not give the same description of the channel, as one can have channels
with the same BER but having different mutual information, and vice
versa. For instance, consider the case $B=0$ corresponding to
$Q_{0|1}=Q_{1|0}=0$ (namely no errors are made in the identification
of the binary symbol at the output). As mentioned before this means
that the information is maximum, i.e. $I(X;Y)=1$.  On the other hand,
for $B=1$ one has $Q_{0|1}=Q_{1|0}=1$, namely the output symbol is
always interpreted in the wrong way: in this case, however, the mutual
information is still unit, and, in fact, it is possible to reconstruct
the transmitted message without ambiguity by swapping the symbols
``0'' and ``1'' at the output. Moreover, one can have different values
of the mutual information that correspond to the same BER, just by
varying the probabilities of error and keeping their sum fixed.

As shown in the above examples, the mutual information gives a more
complete description of the communication channel than the BER, and
hence it should be adopted as the right quantity to characterize the
channel. In the rest of this paper, however, we will use the BER,
because in our case the probabilities of error are so small that BER
is equivalent to the mutual information via Eq. (\ref{equi}).

\section{The quantum communication channel}
\label{qcc}

In this section we specialize the above concepts to the quantum
communication channel, where the information is encoded on quantum
states.  In the first two subsections we give a characterization of
the channel in terms of mutual information, and in terms of
gain and noise figure of the devices inserted along the line.
Finally, in Subsec. \ref{sec:sectfpe} we review the description of
the dynamical evolution of the quantum state along the transmission line,
with the insertion of both amplification and losses.

\subsection{Mutual information in the quantum channel}
\label{miq}

In a quantum channel the symbols to be transmitted are encoded into
density operators $\rho_{x}$ on the Hilbert space ${\cal H}$ of the 
dynamical system that supports the communication. 
The alphabet (which can be either discrete or continuous)
is distributed according to an {\it a priori} probability density $p(x)$,
which also determines the mixture $\bar{\rho}$ given by
\begin{eqnarray}
\bar{\rho}=\int dx\ p(x)\rho_{x} 
\;.\label{u}
\end{eqnarray}
The set of states $\rho_{x}$ and the probability density
$p(x)$ are globally referred to as "encoding".
The transmission line includes all dynamical transformations of the
encoded states between the encoder and the decoder. In the
Schr\"odinger picture these are described by a completely positive
(CP) map $\rho\to \rho_t=E_t (\rho)$ acting on the density matrix
$\rho$ that carries the information through the channel.  The last
step in the communication channel is the decoding operation, where the
recognition of the transmitted symbol is established as the result of
a quantum measurement. In the general case this is conveniently
described by means of a probability operator
measure\cite{helstrom,yuen2} (POM), which extends the notion of
orthogonal-projector-valued measure associated to customary quantum
observables.
The POM $d\^\Pi(y)$ provides
the probability distribution $p[\rho](y)\;dy$ of the random outcome $y$ of
the measurement according to the rule
\begin{eqnarray}
p[\rho](y)\;dy =\hbox{Tr}[\rho\; d\^\Pi(y)]
\;.\label{def-prob}
\end{eqnarray}                                       
In order to have a genuine probability $dP[\rho](y)$, a POM must satisfy
the positivity condition
\begin{eqnarray}
d\^\Pi(y) \ge 0
\label{POM-def2}
\end{eqnarray}
as a consequence of the positivity of the density operator $\^\rho$.
Normalization of the probability $p(y)$ is guaranteed by
the completeness relation 
\begin{eqnarray}
\int d\^\Pi(y)=\^ 1 \;.
\label{POM-def1}
\end{eqnarray}
Each measurement apparatus is described by a POM.
Actually, the POM comes from a customary orthogonal projection-valued 
measure that includes also the degrees of freedom of the apparatus, 
which are partially traced out in order to give a description of
the system alone. Notice that
the correspondence between measuring apparatus and POM's is not 
one-to-one, since different detectors can be described by the same POM.
In a quantum communication channel the POM specifies the conditional 
probabilities in the mutual information (\ref{minf-cont}), namely
\begin{eqnarray}
Q(y|x)\;dy=\hbox{Tr}[E_t(\rho_x)\; d\^\Pi(y)]\;.
\end{eqnarray}
The value of the mutual information depends on all the three steps of
the communication channel: encoding, transmission and decoding.  The
maximum of $I(X,Y)$ over all possible {\it a priori} probability
densities $p(x)$ for a given constraint (typically the average power
along the line) and for an ideal channel (namely $E_t=\iden$ is the identical
transformation) is called ``capacity'' $C$ of the channel, whereas the
global maximum obtained optimizing also over both encoding and
decoding is called ``ultimate quantum capacity''.  For fixed average
power $\bar n$, using the Holevo-Ozawa-Yuen information
bound\cite{OY,Holevo} it can be proved that the ultimate quantum
capacity for a quantum--optical channel is achieved by direct detection
and number-state encoding, with $\bar{\rho}$ being the thermal
state with photon number $\bar{n}$.

\subsection{Gain and noise in amplification}
\label{qa}

Let us now consider the effects of the insertion of an amplifier in a
transmission line. The performances of a quantum amplifier are
described in terms of the gain and the disturbance caused by the
device to the transmitted signal.  As already discussed, the best
quantity to characterize the disturbance is the loss of mutual
information with respect to the case of ideal transmission.  However,
although the noise figure is significant only for Gaussian channels
(see Sec. \ref{gauss}), this is the quantity that is customarily
used to characterize the disturbance of the amplifier.

Both gain and noise figure generally depend on the overall
configuration of the channel, i.e. on the particular encoded states
and on the kind of detection at the decoder.  The gain is defined as
the ratio between the output and the input signals as follows
\begin{eqnarray}
G={{{S}_{out}}\over{{S}_{in}}}\;,
\label{def-G}
\end{eqnarray}
where the signals $S_{in,out}$ are defined as the difference between
expectation values
\begin{eqnarray}
S_{in,out}=\mbox{Tr}\{\^O\ \rho_{in,out}
\}-\mbox{Tr}\{\^O\ \upsilon_{in,out} \},
\label{def-S}
\end{eqnarray}
where $\rho_{out}=E_t(\rho_{in})$,
$\upsilon_{out}=E_t(\upsilon_{in})$, $E_t$ is the CP map of the
amplifier from the input to the output, $\^ O$ is the detected
operator and $\upsilon$ is a reference state. For optical amplifiers
at the input of the channel the state $\upsilon_{in}$ is usually the
vacuum, but it is generally transformed into a non-vacuum state
$\upsilon_{out}$ as it evolves through the amplifier. An amplifier is
called "linear" when the gain does not depend on the input state
$\rho_{in}$. Notice that, as the signal depends on the detection
scheme defined by the observable $\^O$, in principle the linear
character of a device depends on the kind of detection at the output.

The noise figure quantifies the degradation of the signal-to-noise
ratio (SNR) from the input to the output, and is defined as
\begin{eqnarray}
R=SNR_{in}/SNR_{out}\;.
\label{figrum}
\end{eqnarray}
In Eq. (\ref{figrum}) the signal-to-noise ratio is given by
\begin{eqnarray}
SNR={S}^2/{N}\;,
\label{SNR}
\end{eqnarray}
where $N$ is the quantum noise $\<\Delta\^
O^2\>_{\rho}=\hbox{Tr}(\rho\^ O^2)-[\hbox{Tr}(\rho\^ O)]^2$ for
detected observable $\^O$ averaged over the alphabet with {\it a
priori} probability distribution $p(x)$, i.e.
\begin{eqnarray}
N=\int dx\ p(x)\<\Delta\^ O^2_{x}\>_{\rho_{x}} 
\;.\label{N1}
\end{eqnarray}
Notice that the definitions of both signal $S$ and noise $N$ can be
easily generalized to the case of a detector described by a non
orthogonal POM $d\^\Pi$, just replacing the ensemble averages by
\begin{eqnarray}\<y^k\>_p\defi\tr\left[\^\rho\int d\Pi(y)\ y^k\right]
\;.\label{avers}
\end{eqnarray}

When characterizing the quality of an amplifier, usually one resorts
to determining the noise added by the amplifier along the
line. However, a quantitative characterization should be based on the
change of mutual information due to the amplifier insertion.  For this
reason, we define as ``ideal'' an amplifier that does not degrade an
optimized channel, namely that leaves the information invariant in a
channel where the information has been already maximized over all
possible encodings and decodings for a given constraint. As a
rescaling of the probability distribution $Q(y|x)\to Q(G^{-1}y|x)$
leaves the mutual information in Eq. (\ref{minf-cont}) invariant, an
amplifier that rescales the probability by the gain $G$ ---hence maps
the observable $\^O$ to $G\^O$--- is ideal (the pathological situation
of bounded spectrum is not considered, as in this case the concept
itself of amplification is meaningless). For such an ideal amplifier
one has unit noise figure $R=1$, independently of the input state. It
is clear that for a non optimized channel, an amplifier can {\em
increase} the mutual information, and, actually, this is its purpose.
In fact, by rescaling the detection spectrum the amplifier expands the
effective dimension of the Hilbert space that
supports the relevant part of the transmitted information. The loss of
information in a non optimized channel can be caused either by an
irreversible dynamics along the line (this is the case of the loss),
and/or by a mismatch between coding and detection, or by low quantum
efficiency of the detector. In the first case it is clear that the
amplifier must be inserted before the leak of information, namely as a
{\em preamplifier}, whereas in the second case it can be inserted just
before the detector, the quantum efficiency representing essentially an
irreversible evolution inside the detector.  Notice that the
configuration of the channel with the amplifier before the loss may
violate the physical constraint of limited power, and hence the best
configuration is a distributed gain along the line.

In the case of a binary channel, with symbols ``0'' and ``1''
corresponding to the input states $\^\rho_0\equiv\upsilon$ and $\^\rho$
($\upsilon$ being the vacuum), the gain is given by Eq. (\ref{def-G}),
and Eq. (\ref{N1}) rewrites
\begin{eqnarray}
{N}={1\over 2}\left(\langle\Delta\^ O ^2\rangle_{\rho_0}+\langle
\Delta\^ O ^2\rangle_{\rho}\right)\;.\label{N_onoff}
\end{eqnarray}
In Sec. \ref{gauss} we will show the equivalence between BER and
noise figure in the characterization of a Gaussian binary channel,
and in Subsec. \ref{results} the difference between the two
quantities will be revealed in the cases of linear and saturable
amplifiers.  The characterization of an amplifier given in this
section is also suited to describe an attenuator (i.e. a
``deamplifier'') or the loss itself, where now $G<1$, whereas,
generally, for an amplifier one strictly has $G>1$.

\subsection{CP maps, master equation and Fokker - Planck equation.}
\label{sec:sectfpe} 
In this subsection we will briefly introduce the quantum treatment of
the evolution equations for a signal state through a transmission line
in the presence of losses or active devices (e.g. amplifiers).  Both
losses and amplifiers are treated as open systems \cite{Davies}. In an
amplifier the mode of radiation that undergoes amplification---the
"signal" mode---resonantly interacts with other modes of radiation
(parametric amplifier) or with matter degrees of freedom (active
medium amplifier).  The necessary energy for amplification is provided
by ``pump'' modes, whereas ``idler'' modes guarantee resonance
conditions and phase-matching.  All modes different from the signal
mode (i.e. the ``system'') act globally as a reservoir, and the signal
mode undergoes a non unitary irreversible transformation (the
only exception is the ideal phase sensitive amplifier, where the
transformation is unitary).  On the other hand, in the presence of a
loss (a beam splitter or a lossy cavity/fiber) the signal mode
is gradually lost into external modes. The dynamical evolution of the
density matrix $\rho$ of an open system is obtained by partially
tracing over all the external degrees of freedom (globally referred to
as the ``probe'') and is described by a map of the form
\begin{eqnarray}
\rho_t =E_t (\rho)=\mbox{Tr}_P\left[ U_t\, \rho_P\otimes\rho\, 
U_t^{\dagger}\right]
\;,\label{cp1}
\end{eqnarray}
where $U_t$ is a unitary operator and $\rho_P$ is the density matrix for
the probe. 
By expanding the trace over a (numerable) orthonormal set, Eq.
(\ref{cp1}) rewrites as follows
\begin{eqnarray}
E_t(\rho)=\sum_k V_k \rho V_k^{\dagger}\;, \qquad   
\sum_k V_k^{\dagger}V_k =1\;,
\label{cp4}
\end{eqnarray}
which is the Stinespring representation\cite{Davies} of a
normal trace-preserving CP-map.
The infinitesimal version of Eq. (\ref{cp1}) or (\ref{cp4}) 
is usually referred to as the ``master equation''.
Lindblad proved that the most general form of a valid master equation 
is given by\cite{l1}  
\begin{eqnarray}
\de_t\rho_t=\sum_k {\cal D}[A_k]\rho_t\;,\qquad {\cal
D}[A]\rho \doteq A\rho A^{\dagger}-\f{1}{2}\left\{
A^{\dagger}A,\rho\right\} \;,\label{me1}
\end{eqnarray}
where $\{\, ,\, \}$ denotes the anticommutator, $A_k$ are generic
complex operators, and ${\cal D}[A]$ are called Lindblad 
super-operators (a ``superoperator'' is a map that acts on operators
both on the left and on the right sides).

The master equation can be transformed into a differential equation for
an ordinary unknown c-number function (instead of an operator
$\^\rho$) using the Wigner representation of the density matrix.
When the order of derivatives can be truncated at the second one, the
equation has the form of a Fokker Planck equation (FPE)
\begin{eqnarray}
\dt W_s(\al,\alb;t)=\left[ \de_\al\al\ Q_{s}(\al,\alb)+ \de_\alb\alb
Q^*_{s}(\al,\alb)+\frac12\vec{\vec\nabla}_{\al}:\matd_s(\al,\alb)
\right]W_s(\al,\alb;t)
\;,\label{fp1}
\end{eqnarray}
where $(Q_s,Q_s^*)$ is the drift vector, $\matd_s$ the diffusion matrix, $
\vec{\vec\nabla}_{\al}$ is the matrix of second derivatives
\begin{eqnarray}\vec{\vec\nabla}_{\al}\defi
\left(\matrix{\de^2_{\al\alb}&\de^2_{\al\al}
\cr\de^2_{\alb\alb}&\de^2_{\alb\al}}\right)\ ,\end{eqnarray} and
$W_s(\al,\alb)$ is the generalized Wigner function of the radiation
mode with annihilation operator $a$, defined as follows
\begin{eqnarray}
W_s(\alpha,\alpha^*)\doteq
\int {{d^2 \lambda }\over{\pi }}\ 
X_s(\lambda,\lambda^*)\exp\left(\alpha\lambda^* -\alpha^* \lambda
\right) \;.\label{a2:wdef}
\end{eqnarray}
The Wigner function $W_s(\alpha,\alpha^*)$ is the Fourier transform of
the generating function of the $s$-ordered moments
\begin{eqnarray}
X_s(\lambda,\lambda^*) =\hbox{Tr} \left[\^\rho\exp\left(\lambda
a^{\dag}-\lambda^* a \right) \right]\exp\left({1\over 2}s|\lambda |^2
\right)\;,
\label{a2:fcar}              
\end{eqnarray}
with the moments given by
\begin{eqnarray}
\hbox{Tr} \left[\^\rho\left\{ a^{\dag n} a^m \right\}_s \right]=
{{\partial^n }\over{\partial\lambda^n }}
\left. {{\partial^m }\over{\partial (-\lambda^* )^m }}\right|_{\lambda=0}
X_s(\lambda,\lambda^*)=\int {{d^2\alpha}\over\pi}\ W_s(\alpha,\alpha^*)\alpha^{*n} 
\alpha^m\;.
\label{s-ord}
\end{eqnarray}
In the following sections we will only consider the case $s=0$, which
corresponds to symmetrical ordering. Here we notice that the functions
$W_s(\alpha,\alpha^*)$ are not generally probability distributions, as
they are not positive definite, and for this reason they are named
``quasi--probabilities''. Only for $s\leq -1$ one has
$W_s(\alpha,\alpha^*)\geq 0$ for all $\alpha$. In particular for
$s=-1$, which corresponds to antinormal ordering,
$W_s(\alpha,\alpha^*)$ is the probability distribution of the ideal
measurement of the complex field (heterodyne detection), whereas for
$s<-1$ it represents the same measurement with quantum efficiency
$\eta_{het}={2}/({1-s})<1$.

From the Wigner generalized function one can obtain the homodyne
detection probability distribution in form of the marginal integration
\begin{eqnarray}
P_s(x)=\int {{dy}\over{\pi}}W_s(x+iy,x-iy)\;,
\label{marg}
\end{eqnarray}
and now the quantum efficiency $\eta_{hom}$ of the homodyne detector is
related to the ordering parameters by $\eta_{hom}={1}/{(1-s)}$.

\pni In the rest of the present section we will consider a FPE
with constant drift and diffusion coefficients of the form
\begin{eqnarray}
\de_t W_s(\al,\alb)=\left[Q(\de_\al\al+\de_{\al^*}\alb)+
2D_s\de^2_{\al\alb}\right]W_s(\al,\alb).\label{eqfpc}\end{eqnarray}
Eq. (\ref{eqfpc}) can be solved analytically, with solution in form of
the following Gaussian convolution
\begin{eqnarray}
W_s(\al,\alb;t)=\f{1}{\pi\Delta_s^2(t)}
\int d^2\beta\ \exp\left[ -\f{|\al-\beta e^{-Qt}|^2}{\Delta_s^2(t)}\right]
W_s(\beta,\beta^*;0)\;\!,
\label{fp2}
\end{eqnarray}
where
\begin{eqnarray}
\Delta_s^2(t)=\f{D_s}{Q}(1-e^{-2Qt})\;.
\label{vfp2}
\end{eqnarray}
For homodyne detection the marginal integration of the FPE leads to
the Ornstein-Uhlenbeck equation (OUE)
\begin{eqnarray}
\dt P_s(x;t)=\left[Q\dx x+\f{1}{2}D_s\ddx\right]P_s(x;t) 
\;,\label{ou1}
\end{eqnarray}
again with solution in form of the Gaussian convolution
\begin{eqnarray}
P_s(x;t)=\f{1}{\sqrt{2\pi d_s^2(t)}}\int dx'\ 
\exp\left[ -\f{(x-x'e^{-Qt})^2}{2d_s^2(t)}\right]P_s(x';0)\!\;,
\label{ou2}
\end{eqnarray}
where
\begin{eqnarray}
d_s^2(t)=\f{D_s}{2Q}(1-e^{-2Qt})\;.
\label{vou2}
\end{eqnarray}
Notice that the diffusion terms in (\ref{eqfpc}) and (\ref{ou1})
generally depend on the efficiency of the detection apparatus.  It is
straightforward to see from Eqs.(\ref{fp2}) and (\ref{ou2}) that the
gain for the corresponding variable (the field amplitude for
Eq. (\ref{eqfpc}) and the quadrature for Eq. (\ref{ou1})) is given by
$G=e^{-Qt}$, and does not depend on the input state. Therefore, the
processes described by Eqs. (\ref{eqfpc}) and (\ref{ou1}) correspond
to the case of linear amplification for $Q<0$, or to the case of
linear loss for $Q>0$. When the diffusion coefficient in
Eqs. (\ref{eqfpc}) and (\ref{ou1}) vanishes, the evolution of the
corresponding probability distribution is just given by the rescaling
of the field variable $\alpha$, namely
\begin{eqnarray}
W_s(\al,\alb;t)=\frac{1}{G^2}W_s(G^{-1}\al,G^{-1}\alb;0)\;\!
\label{fpr}
\end{eqnarray}
and, for the OUE, for the homodyne variable $x$
\begin{eqnarray}
P_s(x;t)=\frac{1}{G}P_s(G^{-1}x;0)\!\;.
\label{our}
\end{eqnarray}
In this case, for perfect detection (described either by $W_{-1}$ or
by $P_0$) the device is ideal. For detection with $\eta <1$, one can
even have $D_s<0$ \cite{nottingham1}, which means that the ideal
amplifier can improve the SNR and the mutual information. For
heterodyne detection ($W_{-1}$), the ideal amplifier is the so called
phase insensitive amplifier (PIA), which will be described in detail
in Sec. \ref{s:PIA}.  For homodyne detection ($P_0$) the ideal
amplifier is the phase sensitive amplifier (PSA), which is described
by a Ornstein Uhlenbeck equation of the form (\ref{ou1}) with
vanishing diffusion term.  For direct detection, namely detection of
the photon number of the radiation field, the ideal amplifier is
called ``photon number amplifier'' (PNA) \cite{Yuen86}.  In the
following section, we will consider these linear devices inserted in a
linear Gaussian channel with Gaussian input probability.

\section{The linear Gaussian channel}\label{gauss}

A linear Gaussian channel is a particular type of channel where the
evolution equation is given by Eq. (\ref{fp1}) or Eq.  (\ref{ou1}),
i.e. it has a Gaussian Green function, and the encoded states at the
input are themselves described by Gaussian probability distributions. The
time--evolved probability distributions are easily obtained from
Eqs. (\ref{fp2}) and (\ref{ou2}) as follows
\begin{eqnarray}
W_s(\al,\alb;t)=\f{1}{\pi\Delta_s^2(t)}
\exp\left[ -\f{|\al-\al_0e^{-Qt}|^2}{\Delta_s^2(t)}\right]\;\!,\qquad
\end{eqnarray}
\begin{eqnarray}
\Delta_s^2(t)=\f{D_s}{Q}(1-e^{-2Qt})+\Delta_s^2(0)e^{-2Qt}\;,
\label{fp2g}
\end{eqnarray}
and
\begin{eqnarray}
P_s(x;t)=\f{1}{\sqrt{2\pi d_s^2(t)}}
\exp\left[ -\f{(x-x_0e^{-Qt})^2}{2d_s^2(t)}\right]\!\;,\qquad
d_s^2(t)=\f{D_s}{2Q}(1-e^{-2Qt})+d_s^2(0)e^{-2Qt}\;.
\label{ou2g}
\end{eqnarray}
For a Gaussian channel, the probability distribution remains Gaussian
at all times, and the evolution is completely characterized by the
changes of mean value and variance, whereas the drift and diffusion
coefficients are constant.  In this simple case the mutual information
and the noise figure can be easily computed for different
configurations \cite{nottingham1}.

 For the Gaussian channel, we now analyze the relations between the
three different criteria: information, noise, and BER.  
\subsubsection*{Relation between noise and information} 
If we consider a continuous alphabet described by Gaussian states with
variance $\Delta^2$, and transmitted according to an {\it a priori}
Gaussian probability distribution with variance $\delta^2$, the mutual
information takes the form
\begin{eqnarray} I=\frac12\log_2\left(1+\frac{\delta^2}{\Delta^2}\right)
.\end{eqnarray} Hence the mutual information is just a function of the
noise $N=\Delta^2$ and of the variance $\delta^2$ of the {\it a priori}
probability distribution.
\subsubsection*{Relation between noise and BER}  The use of the BER 
actually pertains to the case of a binary channel. For a Gaussian
channel, the two states ``0'' and ``1'' have Gaussian probability
distributions at the output centered around two different values.  The
two Gaussians are overlapping, and one discriminates between ``0'' and
``1'' upon introducing a threshold $\vartheta$, such that one has
``0'' signal for $y<\vartheta$ and ``1'' for $y>\vartheta$.  For
output described by probability densities $q_0$ and $q_1$, the error
probabilities $Q_{0|1}$ and $Q_{1|0}$ are given by
\begin{eqnarray}
Q_{0|1}&=&\int_{x\le \vartheta}dy\ q_1(y)         
\label{fal-al1} \\
Q_{1|0}&=&\int_{x\geq \vartheta}dy\ q_0(y)  \;,
\label{fal-al2}
\end{eqnarray}
 The BER obviously depends on the threshold $\vartheta$, which must be
optimized in order to obtain the minimum value. For Gaussian functions
this corresponds to putting $\vartheta$ at the crossing point between
the graphs of the two output distribution probabilities, and the BER
is just the overlap area. When the input states have the same
variance, the optimal threshold $\vartheta=S/2$ is placed exactly in
the middle between the mean values of the two distributions. From
Eq. (\ref{fal-al1}) one has
\begin{eqnarray} B=1-\Phi\left(\root\of{\frac{SNR}8}\right)\simeq
\root\of{\frac 8{\pi\ SNR}}\ \ e ^{-\frac 18SNR},\end{eqnarray} where
the symbol $\simeq$ stands for asymptotic value for large SNR. As the
error function $\Phi$ is monotone, one can see that for binary Gaussian
channels the SNR and the BER criteria are essentially the same.

For non Gaussian channels, the equivalence between the three different
criteria is generally not valid. In fact, we cannot establish a
general rule for the time evolution of the shapes of the probability
distributions. Therefore, there may occur cases where the
noise figure is small because there is little broadening of the
probability, but the BER is high because the broadening is not
symmetrical around the mean value and the overlap region increases
more than the area below the external tails of the distributions.  As
we will see in Sec. \ref{s:saturable}, this is the case of a
saturable amplifier. In principle, there might be also opposite
situations, where the noise figure gets worse because of a broadening
of the probabilities, but at the same time the BER improves because
the broadening is dominant in the non overlapping region of the two
probabilities.  For non Gaussian channels, therefore, the
characterization of a device performance in terms of noise figure may
be significantly different from the one based on the BER analysis, and
the latter must then be considered instead.

\section{The phase insensitive amplifier}
\label{s:PIA}

A phase insensitive linear amplifier is described by the following master
equation
\begin{eqnarray}
\partial_t\rho_t=2\left[ A{\cal D}[\ad]+B{\cal D}[a]\right]\rho_t\;,\label{mm1}
\end{eqnarray}
where the superoperator $\cal D$ is defined in Eq. (\ref{me1}) and
$A>B$.  As a consequence of the invariance ${\cal
D}[ae^{-i\phi}]={\cal D}[a]$, the device is phase insensitive, namely
it amplifies the field independently of the value of its phase. The
case $B>A$ describes a linear attenuation process.  With $A$ and $B$
proportional to the atomic populations of the upper and lower lasing
levels respectively, Eq. (\ref{mm1}) describes an active medium
amplifier in the linear regime (i.e. far from saturation).  On the
other hand, for $A={{\Gamma}\over 2} \bar m$ and $B={{\Gamma}\over 2}
(\bar m+1)$ the same equation describes a field mode with photon
lifetime $\Gamma ^{-1}$ damped towards the thermal distribution with
$\bar m$ average photons.
 
Eq. (\ref{mm1}) has the following general CP-map solution\cite{liouv}
\begin{eqnarray}
\rho_t=\mbox{Tr}_P[U_t\,\rho\otimes\nu\, U_t^{\dag}]\;,\qquad
U_t=\left\{\begin {array}{ll}
\exp[-\arctan\sqrt{e^{\Gamma t}-1}(a b^\dagger-a^\dagger b)]
& (B>A)\;,\\
\exp[-\hbox{arctanh}\sqrt{1-e^{-\Gamma t}}(a^\dagger 
b^\dagger-a b)]
& (A>B)\;,\end{array}
\right.
\label{rr1}
\end{eqnarray}
where $\Gamma/2=|A-B|$, and the second field mode $b$ is an additional
mode, called ``idler'', that is initially in the thermal state $\nu$
with average photons $\bar m=\mbox{min}\{A,B\}/|A-B|$ [$A$ and $B$
must be non negative, otherwise one would have negative idler
photons].  The idler mode $b$ is needed for the unitarity requirement
of the quantum evolution of the fields \cite{ijmp}. Ideal
amplification is achieved when the idler mode $b$ is initially in the
vacuum state. In the case of finite temperature, when the mode $b$ is
initially in the thermal state with a non vanishing number of photons
$\bar m$, the amplifier is not ideal ($B\ne 0$).

Let us now study the performance of the PIA for different kinds of
detection. Corresponding to the master equation (\ref{mm1}) one can
obtain a Fokker-Planck equation of the form (\ref{fp1}) with $Q=B-A$
and $2D_s=A+B+s(A-B)$. For ideal heterodyne detection ($s=-1$) the
device is ideal for $B=0$ (ideal phase insensitive amplifier), because
the diffusion coefficient vanishes. This corresponds, for example, to
the case of complete inversion between the two lasing levels in an
active medium amplifier. On the contrary, linear attenuation is always
non ideal for heterodyne detection.  The Ornstein Uhlenbeck equation
has the form (\ref{ou1}), and for ideal homodyne detection ($s=0$) the
PIA is never ideal.  To study ideal direct detection, where the photon
number operator is measured, we need the evolution of the number-state
probability.  From Eq. (\ref{rr1}) for $A>B$, and for input number
state $|m\>\< m|$, we have \begin{eqnarray}
P_t(n)\equiv\<n|\^\rho_t|n\>=\left\{\matrix{ \frac{(e^{2\Gamma
t}-1)^{n-m}}{e^{2(n+1)\Gamma t}}\ \frac{n!}{(n-m)!m!}\qquad
&\mbox{for\ } n\geq m\cr 0\qquad &\mbox{for\ }
n<m}\right.,\label{probnum}\end{eqnarray} whereas in the case of input
coherent state $|\al\>\<\al|$ we have \begin{eqnarray}
P_t(n)=\sum_{h=0}^n\frac{(e^{2\Gamma t}-1)^h} {e^{2(n+1)\Gamma t}}\
\frac{n!}{h![(n-h)!]^2} e^{-|\al|^2} |\al|^{2(n-h)}.\end{eqnarray}
These results show that the PIA is not ideal for direct detection,
since the output probability distribution is not simply rescaled. This
is the cause of the unavoidable noise of quantum nature that the PIA
introduces for direct detection. Moreover, from Eq.  (\ref{probnum}),
we see that for input number states a non-vanishing BER is developed
during the time evolution starting from initial BER $B=0$. Figure
\ref{f:piaber} illustrates the mechanism of BER occurrence for
coherent state input.

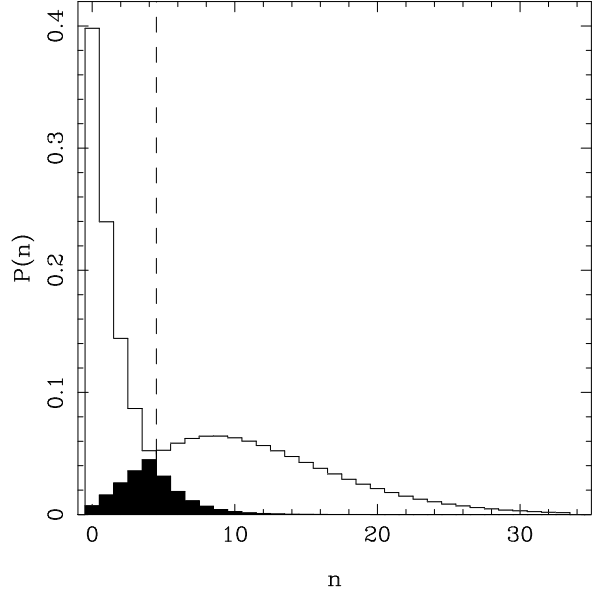
\autofig{hbt}{.5}{piacal} {The picture shows the BER (shaded surface)
at the output of a PIA. The left histogram is the output distribution
corresponding to input bit ``0'' (i.e. the vacuum); the right
histogram is the output distribution when the input is bit ``1'' (a
coherent state, in this case).  The vertical line indicates the value
of the threshold $\vartheta$ that optimizes the BER.  }{f:piaber}

We'll now analyze in more detail the role of the added noise. The
output mode of the electromagnetic field is \begin{eqnarray}
a_{out}=U^\dagger a_{in}U,\end{eqnarray} where $U$ is defined in
Eq. (\ref{rr1}). The linear transformation between the fields is given
by \begin{eqnarray}
a_{out}=G^{1/2}a_{in}+(G-1)^{1/2}b_{in}^\dagger.\end{eqnarray} Thus,
in the case of direct detection, the output average number of photons
of the PIA is \begin{eqnarray}\<(a^\dagger a)_{out}\>=G\<(a^\dagger
a)_{in}\>+(G-1)(\<(b^\dagger b)_{in}\>+1), \end{eqnarray} where
$G=e^{\Gamma t}$ is the amplifier gain.  The idler mode $b$ is
responsible for a non vacuum output when there is no input ({\it
spontaneous emission}), i.e. with the mode $a_{in}$ in the vacuum
state.  The noise figure is obtained from the output noise defined
in Eq. (\ref{N_onoff}). For binary channels and under the condition of
vacuum state for the bit ``0'', the output noise is \cite{dm}
\begin{eqnarray} N_{out}&=&\frac 12\Bigl[{G}\<\^n_a\>_{
\scriptsize{\mbox{bit 1}}}+
2({G}-1)\<\^n_b+1\>+2{G}({G}-1)\<\^n_b+1\>\<\^n_a\>_{\scriptsize{\mbox{bit\
1}}}+\cr &+&2({G}-1)^2\<2\^n_b+1\>+ {G}^2\<\^n_a\>_{
\scriptsize{\mbox{bit\ 1}}}({\cal F}_a -1)+2(G-1)^2\<\^n_b\>({\cal
F}_b-1)+\cr &+&2G(G-1)\Bigl(\<b{^\dagger}^2\>\<a{^\dagger}^2\>+
\<b^2\>\<a^2\>\Bigr)\Bigr]\label{noisef}\end{eqnarray} where
$\^n_a\equiv a^\dagger a$, $\^n_b\equiv b{^\dagger}b$ and ${\cal F}$ is the
Fano factor of the distribution, i.e.
\begin{eqnarray}{\cal F}=\frac{\<\Delta \^n^2\>}{\<\^n\>}.\end{eqnarray} 
The various contributions to the noise are usually referred to as
[following the same order in Eq. (\ref{noisef})]:
(i) quantum fluctuations of the amplified signal; (ii) amplified
parametric spontaneous emission; (iii) quantum beat between the first
and the second terms; (iv) self-beat of parametric spontaneous emission;
(v) excess noise of the signal and (vi) of the idler; (vii) coherent terms.
It is easy to see that for strong input signals $\<\^n_a\>\gg 1$,
high gain $G\gg 1$, and vacuum idler state, the noise figure is 
\begin{eqnarray}{R}\simeq\frac{2\<\^n_a\>}{\<\Delta \^n_a^2\>}.\end{eqnarray} 
For coherent input states, the Fano factor is ${\cal F}=1$, and the
noise figure is \begin{eqnarray}{R}\simeq 2\simeq 3\mbox{\
dB}.\end{eqnarray} This is the minimal noise figure that can be
achieved by a phase insensitive quantum linear amplifier for coherent
input states and under the above assumptions: it is usually referred
to as {\it Standard Quantum Limit} (SQL).  As we have shown, and as
shown in Ref. \cite{saturation}, this limit is not unsurpassable, but
it is just a peculiarity of the linear character of the PIA.

\section{The saturable laser amplifier}\label{s:saturable}
As we have seen in the previous section, the PIA is not ideal for
direct detection.  In this case the ideal amplifier is the PNA
proposed by Yuen \cite{Yuen86}.  Even though the Hamiltonian for the
PNA has been derived (see Refs. \cite{darianopna} and \cite{ijmp}),
its practical realization is unknown, and a way to approach such an
ideal device by concrete amplifiers still remains an open problem. In
this section we investigate the possibility of approximating the
behavior of the PNA by means of a laser amplifier working in a
non-linear regime.  The laser will be studied on the basis of the
theory of Haake and Lewenstein \cite{haake}.  The underlying master
\eq\ describes uncorrelated two level atoms interacting with a
radiation mode in a cavity. The evolution of radiation in the laser is
described by means of a FPE that is derived from the \meq\ after
adiabatic elimination of the atomic variables. In this
section we briefly recall the Haake--Lewenstein theory, giving the
detailed FPE and a discussion of the validity limits of the theory.

\subsection{The \meq .}
The model describes $N$ two level atoms interacting with a single mode
of the electromagnetic field (dipole interaction in the rotating wave
approximation), whereas the non--lasing lossy modes and the pump
mechanisms are taken into account in form of baths. The complete \meq\
that describes atoms and radiation is given by
\begin{eqnarray}\de_t\^R(t) =L\^R(t),\label{meq}\end{eqnarray} where
$\^R$ denotes the joint atom--radiation density matrix, and $L$ is the
Liouvillian \begin{eqnarray}
L=L_a+L_f+L_{af},\label{liouv}\end{eqnarray} where the atomic
contribution $L_a$ is
\begin{eqnarray}L_a=\sum_{j=1}^NL_j\ ;\ {L}_j=\frac{\gamma_\|}{2}(1+\sigma_0){\cal
D}[\^{\sigma_+}_j] + \frac{\gamma_\|}{2}(1-\sigma_0){\cal
D}[\^{\sigma_-}_j] + \frac{1}{2} \biggl(\gamma_\perp-
\frac{\gamma_\|}{2}\biggr){\cal D}[\^{\sigma_z}_j],\end{eqnarray}  
the field contribution $L_f$ is given by
\begin{eqnarray}L_f=\gamma   n_{th}{\cal  D}[\^a{^\dagger}]+\gamma(n_{th}+1){\cal D}[\^a],\end{eqnarray} 
and the interaction term $L_{af}$ is 
\begin{eqnarray}{L}_{af}\^R(t)= g[\^S_+\^a-\^S_-\^a^+,\^R(t)].\end{eqnarray} 
The quantities involved in the above \eq s have the following meaning:
 $\sigma_0$ is the stationary expectation value for the population
 inversion $\^\sigma_z$ without field (non--saturated inversion);
 $\gamma_\perp$ is the decay rate of the atomic polarization;
 $\gamma_\|$ the decay rate of the population inversion; $\cal D$
 is the Lindblad superoperator defined in \eq\ (\ref{me1}); $\^\sigma_+$, 
$\^\sigma_-$ and $\^\sigma_z$ are the Pauli operators

\begin{eqnarray}\^\sigma_+=\left(\matrix{0&1\cr0&0\cr}\right),\ \
\^\sigma_-=\left(\matrix{0&0\cr1&0\cr}\right),\ \
\^\sigma_z=\left(\matrix{1&0\cr0&-1\cr}\right)\; ;\end{eqnarray} $\^a$
is the field annihilation operator; $\gamma$ is the decay rate of the
cavity; $n_{th}$ is the mean number of thermal photons in the cavity;
$\^S_\pm=\sum_{j=1}^N \^{\sigma_\pm}_j$; $g$ is the electrical dipole
coupling between atoms and radiation.

\subsection{Adiabatic expansion of the master equation}
\label{aeme}
From the \meq\ (\ref{meq}) one proceeds to eliminate the fast atomic
degrees of freedom by means of an adiabatic expansion.  The
Liouvillian is separated into two parts
\begin{eqnarray}L=L_0+\Delta
L,\label{liouvi}\end{eqnarray} where $L_0$ is the zero order ``fast'' term and
$\Delta L$ is the ``slow'' remaining Liouvillian.  The zero order
Liouvillian $L_0$ drives the fast (atomic) variables toward
equilibrium with the slow one (the lasing mode) with a characteristic
decay time $t_*$.  Any initial state $\^R(0)$ is evolved by $L_0$ into
a state $\^R_*$, after a time $t\gg t_*$, such that $L_0\^R_*=0.$ The
spirit of the adiabatic approximation is to consider the dynamical
evolution on a time scale $t\gg t_*$.  Eventually, one obtains a
second order expansion for the master \eq\ of the slow variables in
the form
\begin{eqnarray}\de_t\^\rho(t)=\Bigl\{\tr_0\Bigl[\Delta L\^A+\int^t_0 d\tau'\
\Delta L (e^{L_0\tau'}-\^A\ \mbox{Tr}_0)\Delta
L\^A\Bigr]\Bigr\}\^\rho(t),\end{eqnarray} where
$\^\rho(t)\defi\tr_0[\^R(t)]$ is the reduced slow density matrix
describing the lasing mode alone, $\tr_0$ denotes the partial trace
over the fast (atomic) degrees of freedom, and
\begin{eqnarray}\^A= \^R_*\Bigl(\tr_0[\^R(0)]\Bigr)^{-1}\end{eqnarray}
is the atomic state that adiabatically follows the field.

\subsection{The Fokker - Planck \eq}
\label{FPE}
Using the Wigner function representation derived in Subsec.
 \ref{sec:sectfpe} the annihilation and creation operators become
 differential operators, and the \meq\ is written in form of a
 FPE. One introduces the Wigner function
\begin{eqnarray}\^W(\alpha,\alpha^*) \defi \int\frac{d^2\lambda}{\pi^2}
e^{\alpha\lambda^*-\alpha^*\lambda} \mbox{Tr}_f[\^R(t)e^{\lambda
\^a^+-\lambda^* \^a}],\label{wigeq}\end{eqnarray} where $\tr_f$
denotes the partial trace over the field and hence $W$ is still an
atomic operator. In the Wigner representation the Liouvillian rewrites
\begin{eqnarray}{L}_fW=\frac{\gamma}{2}\{\de_\alpha\alpha+\de_{\alpha^*}\alpha^* +
(2n_{th}+1) \de^2_{\alpha\alpha^*}\}W\end{eqnarray}
\begin{eqnarray}{L}_{af}W=g\biggl[\alpha\^S_+-\alpha^*\^S_-,W\biggr]+\frac g2
\de_\alpha\biggl( W\^S_-+\^S_-W\biggr )+\frac
g2\de_{\alpha^*}\biggl(\^S_+W+W\^S_+ \biggr),\end{eqnarray} whereas
$L_a$ is unchanged. In the following we will consider zero thermal
photons ($n_{th}=0$) in the lasing mode.

\riga\par We are now ready to perform the adiabatic expansion of the
master \eq\ described in Sec. \ref{aeme}, and the Wigner
representation (\ref{wigeq}) makes this task much easier.  We start
from separating the total Liouvillian $L\defi L_a+L_f+L_{af}$ into the
two parts $L_0$ and $\Delta L$ in Eq. (\ref{liouvi}). Haake and
Lewenstein's procedure considers
\begin{eqnarray} L_0W \defi {L}_a+g[\alpha
\^S_+-\alpha^*\^S_-,W]\label{defl0}\end{eqnarray}
\begin{eqnarray}\Delta L \defi {L}_f +\frac g2\de_\alpha\biggl(W
\^S_-+\^S_-W\biggr)+\frac g2\de_{\alpha^*}\biggl(\^S_+W +W
\^S_+\biggr).\label{deltal}\end{eqnarray} The expansion term $\Delta
L$ contains the only parts of the Liouvillian that are responsible for
the radiation evolution.  \par Two hypotheses are needed to assure the
validity of the theory.  The photon lifetime in the cavity must be
longer than the atomic decay rates \begin{eqnarray}\gamma \gg
\gpar,\gort.\end{eqnarray} Such condition is consistent
 with the adiabatic
expansion, since the atoms must have a characteristic time slower than
the radiation one.  Moreover, the time scale at which we consider the
solution of the FPE must be bigger than the time $t_*$ at which the
atomic variables are traced out, namely
\begin{eqnarray}t\gg t_*
\sim\gamma_\|^{-1},\gamma_\perp^{-1}.\label{tstar}
\end{eqnarray} This condition
guarantees that the atomic variables have reached the equilibrium with the
field variables. 

\par Instead of the field variable $\al$ it is convenient to use the
variable $u\defi\al/{\root\of{n_s}}={2\al g}/{\root\of{\gpar\gort}}$,
where the field is rescaled by the number of saturating photons
$n_s\defi\frac{\gpar\gort}{4g^2}$.  \2righe \par\noindent With the
procedure outlined above, in Ref.\cite{haake} the following FPE is obtained
\begin{eqnarray}\de_tW(u,u^*)={\cal L}W(u,u^*),\label{eqfp}\end{eqnarray}
where now $W$ is the usual c--number Wigner function of the lasing
mode and the Liouvillian is the differential operator \begin{eqnarray}{\cal L}\defi
\de_uu\ Q_u+\de_{u^*}u^*Q^*_u+\de^2_{uu}D_{uu}
+\de^2_{{u^*}u^*}D^*_{uu}+2\de^2_{uu^*}D_{uu^*}.\label{lio}\end{eqnarray}
The drift coefficients and the diffusion matrix in \eq\ (\ref{lio})
are given by
\begin{eqnarray}Q_u &\defi& 
\frac\gamma 2 \Biggl\{1-\frac{2\sigma_0C}{1+|u|^2}+ \frac{\sigma_0N}{2n_s(\
1+|u|^2)^3}[(1+f)|u|^2 -f] \cr
&+&\frac{\sigma_0^2Cf}{n_s(1+|u|^2)^4}\left[N(1-|u|^2)-2|u|^2\right]\cr &+&
\frac C{2n_s(1+|u|^2)^4}
\left[-2\sigma_0^2N|u|^2+\sigma_0^2(1-|u|^2)+(3+|u|^2)(1+|u|^2)^2\right]
\label{drift}\end{eqnarray}
\begin{eqnarray}
D_{uu}\defi-\frac{C\gamma u^2}{4n_s(1+|u|^2)^3}[\sigma_0^2(1+2f)+(1+|u|^2)^2]
\label{driftaa}\end{eqnarray}
\begin{eqnarray}
D_{uu^*}\defi\frac\gamma 4\Biggl[ \frac 1{n_s}+\frac
C{n_s(1+|u|^2)^3}[(1+|u|^2)^2(2 +|u|^2)-|u|^2\sigma_0^2
(1+2f)]\Biggr]\;,\label{driftab}\end{eqnarray} with
$f\defi\frac\gpar{2\gort}$ and $C\defi\frac{g^2N}{\gamma\gort}$
denoting the cooperation parameter of the laser.  \par\noindent The
independent parameters that appear in the Fokker--Planck \eq\ are six:
the cooperation parameter $C$, the non--saturated population inversion
$\sigma_0$, the number of lasing atoms $N$, the decay rate of the
optical cavity $\gamma$, the ratio between the atomic decay rates $f$,
the number of saturating photons $n_s$.  For the validity of the FPE
we remember that these parameters must satisfy the following
conditions:
\begin{enumerate}\item For the adiabatic approximation one has
${\gamma_\|},\gamma_\perp\gg \gamma,$ which implies that
\begin{eqnarray}\frac N{n_s}\ll 4C,\ \frac {Nf}{n_s}\ll
2C.\end{eqnarray}
\item For the trace time $t_*$ condition one has $t\gg t_*\ \sim
\gamma_\|^{-1},\ \gamma_\perp^{-1}$, which implies that
\begin{eqnarray}\frac 1{\gamma
t}\ll\frac{4n_sC}{N},\frac{2n_sC}{fN},\end{eqnarray} where $t$ is the
time scale at which we consider the solution of the FPE.
\item For the validity of the adiabatic expansion
($\frac{g^2}{\gpar\gort}\ll 1$) one needs large saturation numbers
\begin{eqnarray}n_s\gg \frac 14.\end{eqnarray}
\end{enumerate}
\riga With the above theoretical description of the laser we can now
 study the performance of a laser amplifier used as a traveling wave
 optical amplifier (TWOA), where both cavity mirrors are
 semitransparent with high transmissivity: one mirror lets the input
 radiation enter the cavity, while the other lets the amplified
 radiation exit. In Subsec. \ref{results} we will describe the TWOA by
 the laser FPE (\ref{eqfp}), with the initial Wigner function
 corresponding to the input state of radiation.

\section{Monte Carlo simulations: numerical results}
\label{checks}
We have studied the laser FPE \eq\ (\ref{eqfp}) numerically using the
Monte Carlo simulation method originally proposed in
Ref. \cite{moroni}. In this section we present the numerical results
from simulations and compare the laser amplifier with the ideal PIA in
terms of noise and bit error rate.
\subsection{Some tests of the theory and of the simulation method}
Ref. \cite{moroni} includes extensive checks of the Monte Carlo
simulation method on the basis of analytical models. Additional checks
to tune the numerical parameters (such as the integration time step)
for a specific model can be performed by comparing the simulation
results for stationary quantities with those from the method of the
pseudopotential. Also, very easily, one can check results in the
linear regime against the PIA. Regarding the validity limits of the
FPE (\ref{eqfp}) and the underlying adiabatic approximation, a careful
test is in order, also in consideration of alternative adiabatic
approximations that lead to different Fokker-Planck \eq s, as, for
example, in the theory by Lugiato, Casagrande and Pizzuto
\cite{lugiato} that was considered in Ref. \cite{saturation}. In the
derivation of Ref. \cite{lugiato}, differently from Ref. \cite{haake},
the zeroth order part of the Liouvillian $L_0$ contains $L_a$ and
$L_f$, whereas $\Delta L=L_{af}$ corresponds to a perturbation
expansion in $g$, rather than an adiabatic approximation. As a result,
the FPE has the same diffusion matrix, but the drift lacks the second
order contribution in the adiabatic expansion parameters. A test of
the goodness of the FPE can be done by comparing results versus those
from the solution of the \meq\ (\ref{meq}) without the adiabatic
approximation, evaluating the partial trace over the atomic Hilbert
space after the time $t_*$ in Eq. (\ref{tstar}). The \meq , in turn,
can be solved using the Quantum Jump (QJ) method \cite{molmer,dum},
and in this way one has a completely alternative simulation for the
laser and a very stringent test for the FPE (\ref{eqfp}).

\autofig{htb}{.5}{confr} 
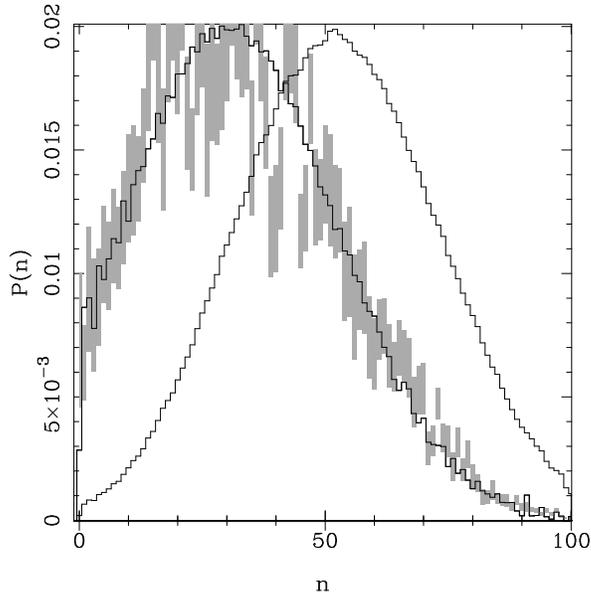
{The output photon number distribution for a
laser. The histogram with the error bars comes from a Quantum Jump
simulation of a one atom laser. The other two histograms are obtained
from the theories of Haake {\it et al.} \cite{haake} (histogram on
the left) and Lugiato {\it et al.}  \cite{lugiato} (histogram on the
right). Note the agreement of the theory of Haake 
{\it et al.} with the numerical
solution of the \meq . The laser parameters used in the simulation are
$C=30;\ f=1;\ n_s=15;\ N=1;\ \sigma_0=0.05;\ \gamma t=\infty$.
}{f:confrlughaa} In Fig. \ref{f:confrlughaa} we report a sample test
with the detailed number probability distribution for the stationary
state from the QJ method for a one-atom laser versus the results of
simulations of the FPE of both Refs. \cite{haake} and
\cite{lugiato}. One can see that in this regime the FPE of
Ref. \cite{haake} perfectly reproduces results from the QJ, whereas
there are discrepancies with results from the FPE of
Ref. \cite{lugiato}. We have analyzed a wide range of parameters, and
found always perfect agreement for the FPE (\ref{eqfp}), within the
validity limits given in Subsec. \ref{FPE}.

\subsection{Comparison between linear and saturable amplifiers}
\label{results}
\begin{figure}[hbt]
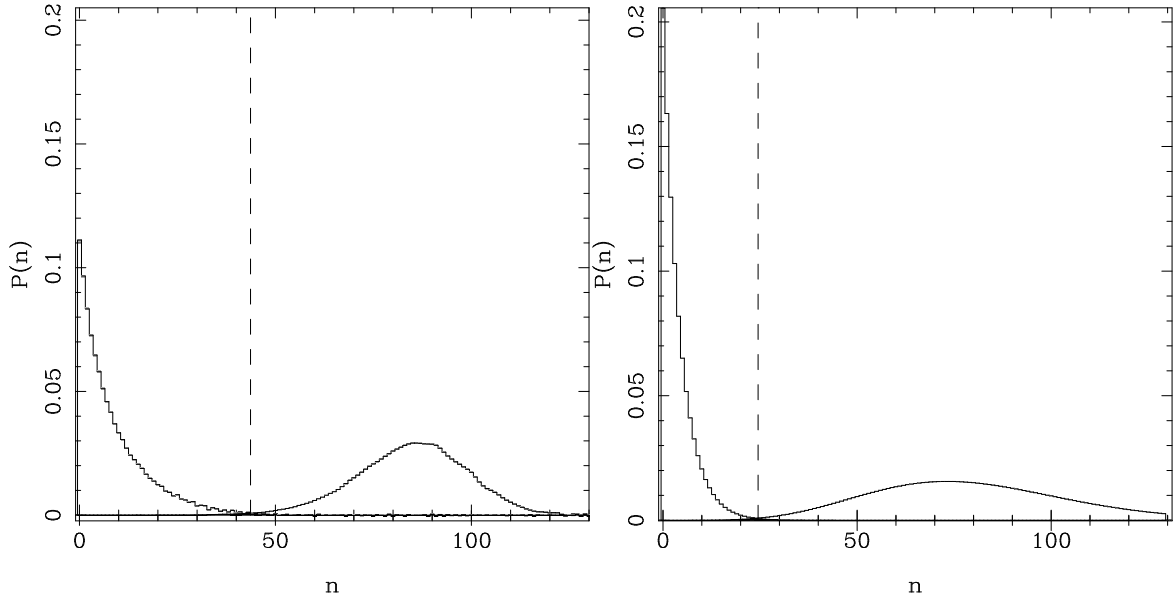
\begin{center}
\getFIG{.5}{ber}\getFIG{.5}{berpia} 
\end{center}
\caption[fake] {\footnotesize Comparison between the output photon
number distributions $p_n$ of the saturable amplifier (left figure) and ideal
PIA (right figure). The parameters used in this simulation are
$C=4.5 ;\ n_s=N=55 ;\ \sigma_0=1;\ \gamma t=0.2 ;\ |\alpha|=3.95 $
(amplitude of the input coherent state pertaining bit ``1'').  This
amplifier has a gain of $6.873$ dB.
\label{f:pia.vs.sat}}\end{figure}

In this subsection we compare the ideal (i.e. with vacuum idler) PIA
and the laser saturable amplifier described by the FPE
(\ref{eqfp}). The amplifier is inserted in a binary communication
channel, with the ``0'' bit encoded on the vacuum state $|0\>$ and the
``1'' bit encoded on a fixed coherent state $|\al\>$. We analyze the
gain $G$, the noise figure $R$ and the bit error rate $B$. In the
linear regime the laser is equivalent to a PIA with gain
\begin{eqnarray}
G=\exp[2\gamma t(1-2\sigma_0C)] 
\;,\label{piagain}\end{eqnarray} but with a non vacuum
idler mode with number of photons
\begin{eqnarray}
n_b=\frac {\mbox{min}\left\{C(1+\sigma_0),C(1-\sigma_0)+1\right\}}
{|2C\sigma_0-1|}.\end{eqnarray} From Eq. (\ref{piagain}) one can
see that the laser amplifies the radiation if the threshold condition
$\sigma_0>1/2C$ is satisfied.  The idler photons produce additional
noise (see Eq. (\ref{noisef})) and the ideal PIA has always a
better noise figure than the laser. Correspondingly, from numerical
simulations we found that the ideal PIA also performs better than the
laser in terms of BER. In the saturation regime, on the other hand,
the noise figure of the laser can be much smaller than the 3-dB
standard noise figure of the PIA, as assessed in
Refs. \cite{saturation},\cite{footnote}. However, in such a regime,
the noise figure $R$ is not the significant criterion of goodness to
be considered, as explained in Sec. \ref{gauss} and pointed out in
Refs. \cite{comment} and \cite{reply}. As a matter of fact, the
numerical analysis shows that the BER for the laser is always worse
than (or, at best, very close to) that of the PIA, even if the
noise figure for the laser is very small. The mechanism leading to
such discrepancy
between noise figure and BER criteria is illustrated in Fig.
\ref{f:pia.vs.sat}, where the output
probability distribution of the two amplifiers is plotted in a regime
of strong saturation for the laser.

 The saturated laser tends to cut off the right tail of the output
distributions for high photon numbers. On the other hand, the ideal
PIA, being a linear device, evolves both tails of the distribution in the
same way.  If we analyze the two amplifiers at the same gain, the left
tail of the bit ``1'' distribution will be higher for the saturable laser
than for the ideal PIA, since saturation needs a higher contribution
from the left tail in order to achieve the same gain, and the left
tail is less influenced by saturation than the right one. For the same
reason the bit ``0'' distribution at the output of the laser turns out to
be much more broadened than for the PIA. As a consequence the overlap
region of the two (``0'' and ``1'') distributions -- i.e. the bit
error rate -- is higher in the saturable laser than in the PIA.

\section{Conclusions.}\label{conclusions}
In this paper we have shown that, for direct detection, saturation
effects in a traveling wave laser amplifier cannot yield BER better
than those for a PIA, although they greatly reduce the noise figure
greatly below the SQL of 3 dB. In fact, the saturation mechanism
broadens the output probability distributions asymmetrically, in such
a way that the noise figure is improved, whereas the overlap between
the distributions, which gives the BER, is increased. This mechanism
has been shown on the basis of a careful Monte Carlo simulation of a
traveling wave laser amplifier. The question now is: is it possible
to improve the BER of the PIA in some way? The answer is clearly
affirmative, if one just considers that at least the ideal PNA could
in principle achieve zero BER when amplifying input number states, or
has $B=e^{-|\al|^2}$ independent of the gain $G$ for input bit ``1''
in the coherent state $|\al\>$, a value anyway much smaller than the
corresponding BER for the PIA. Now the problem is how to achieve or to
approach a PNA by a real device, and from the present analysis we
conclude that the saturated laser does not approximate a PNA better
than a PIA, even though it has a very low noise figure.

\Begref
\bib{OY} H. P. Yuen and M. Ozawa, Phys. Rev. Lett. {\bf 70}, 363 (1993).
\bib{3db} For example see: J.A. Levenson, I. Abram, Th. Rivera and Ph. 
Grangier, J. Opt. Soc. Am. B {\bf 10}, 2233 (1993).
\bib{saturation} G. M. D'Ariano, C. Macchiavello, M. G. A. Paris, Phys.
Rev. Lett. {\bf 73}, 3187 (1994).
\bib{Yuen86} H. P. Yuen, Phys. Rev. Lett. {\bf 56}, 2176 (1986).
\bib{comment} O. Nilsson, A. Karlsson, J.-P. Poizat, E. Berglind, Phys.
Rev. Lett. {\bf 76}, 1972 (1996).
\bib{reply} G. M. D'Ariano, C. Macchiavello, M. G. A. Paris, Phys.
Rev. Lett. {\bf 76}, 1973 (1996).
\bib{haake} F. Haake, M. Lewenstein, Phys. Rev. A {\bf 27} 1013 (1983).
\bib{helstrom} C. W. Helstrom, {\em Quantum Detection and Estimation 
Theory} (Academic Press, New York, 1976).
\bib{yuen2} H. P. Yuen, Phys. Lett. {\bf 91A}, 101 (1982).
\bib{Holevo} A. S. Holevo, Probl. Inf. Transm. {\bf 9}, 177 (1973).
\bib{nottingham1} G. M. D'Ariano, C. Macchiavello and M. G. A. Paris,
{\it Information gain in quantum communication channels},
in ``Quantum Communications and Measurement'', ed. by 
V. P. Belavkin et al., Plenum Press, New York (1995), p.339.  
\bib{Davies} E.~B. Davies, {\em Quantum theory of open systems}, 
(Academic Press, London, New York, 1976).
\bib{l1} G. Lindblad, Commun. Math. Phys. {\bf 48}, 119 (1976).
\bib{liouv} G. M. D'Ariano, Phys. Lett. A {\bf 187}, 231 (1994).
\bib{lugiato}L. A. Lugiato, F. Casagrande and L. Pizzuto,
Phys. Rev. A {\bf 26} 3438 (1982). 
\bib{molmer} K. M\o lmer, Y. Castin, J. Dalibard, J. Opt. Soc. Am. B {\bf 10}, 524 (1993)
\bib{dum}R.Dum, P.Zoller and H.Ritsch, Phys. Rev. A {\bf 45}, 4879 (1992).
\bib{darianopna} G. M. D'Ariano, Phys. Rev. A {\bf 45}, 3224 (1992)
\bib{moroni} G. M. D'Ariano, C. Macchiavello, S. Moroni, Mod. Phys. Lett. B
{\bf 8}, 239 (1994).
\bib{footnote} For the parameter choice exploited in Ref. \cite{saturation}
 both theories of Haake \cite{haake} and Lugiato \cite{lugiato} give the same
results.
\bib{ijmp} G. M. D'Ariano, Int. J. Mod. Phys. B {\bf 6}, 1291 (1992)
\bib{dm} G. M. D'Ariano, C. Macchiavello, Phys. Rev. A {\bf 48}, 3947 (1993)
\Endref
\end{document}